\documentclass[iop,pra,eqseqnum,nofootinbib,amsmath,amssymb]{revtex4-1}

\def\a{\hat a}
\def\ad{\hat a^\dagger}
\def\dxx{d^\times\!x}
\def\eq#1{eq.$\,$(\ref{#1})}
\def\Eq#1{Eq.$\,$(\ref{#1})}
\def\eqs#1{eqs.$\,$(\ref{#1})}

\def\half{{\textstyle{\frac12}}}
\def\halfs{{\textstyle\frac{s}{2}}}
\def\Hbk{\hat H_{\rm\sss BK}}
\def\Hho{\hat H_{\rm\sss HO}}
\def\la{\langle}
\def\p{\hat p}
\def\R{{\mathbb R}}

\def\ra{\rangle}
\def\Re{\mathop{\rm Re}}
\def\Q{{\mathbb Q}}
\def\Qp{{{\mathbb Q}_p}}
\def\Qinf{{{\mathbb Q}_\infty}}
\def\Qpx{{{\mathbb Q}_p^\times}}
\def\Qv{{{\mathbb Q}_v}}
\def\Qvx{{{\mathbb Q}_v^\times}}
\def\sgn{\mathop{\rm sgn}}
\def\sss{\scriptscriptstyle}

\def\Ubk{\hat U_{\rm\sss BK}}
\def\Uho{\hat U_{\rm\sss HO}}
\def\x{\hat x}
\def\xt{\tilde x}
\def\Z{{\mathbb Z}}

\begin{document}

\title{The Berry--Keating Hamiltonian \\ and the Local Riemann Hypothesis}

\author{Mark Srednicki}
\email{mark@physics.ucsb.edu}

\affiliation{Department of Physics, University of California, Santa Barbara, CA 93106}

\date{\today}

\begin{abstract} 
The local Riemann hypothesis states that the zeros of the Mellin transform of a harmonic-oscillator eigenfunction (on a real or $p$-adic configuration space) have real part 1/2.  For the real case, we show that the imaginary parts of these zeros are the eigenvalues of the Berry--Keating hamiltonian
$\Hbk=(\x\p+\p\x)/2$ projected onto the subspace of oscillator eigenfunctions of lower level.  
This gives a spectral proof of the local Riemann hypothesis for the reals,
in the spirit of the Hilbert--P\'olya conjecture. 
The $p$-adic case is also discussed.  
\end{abstract}

\maketitle

\section{Introduction}

The Riemann hypothesis states that all ``nontrivial" zeros (those with real part between zero and one) of the Riemann zeta function $\zeta(s)$ have real part equal to exactly one half \cite{Edw}. 
One approach to proving the Riemann hypothesis is via the Hilbert--P\'olya conjecture 
(for a historical review see \cite{HP})
which posits that the imaginary parts of the nontrivial zeros are the eigenvalues
of a self-adjoint hamiltonian operator $\hat H$.  
A proof of the Riemann hypothesis would then follow from
showing that the spectral determinant of $\hat H$ takes the form
\begin{equation}
\det(E-\hat H) = B(E)\Lambda(\half+iE)\;,
\label{det}
\end{equation}
where $B(E)$ is a normalizing factor with no zeros, $\Lambda(s)$ is the ``completed'' zeta function,
\begin{equation}
\Lambda(s) :=  \pi^{-s/2}\Gamma(\halfs)\zeta(s)\;,
\label{Lam}
\end{equation}
and the zeta function itself is defined, for $\Re s >1$, by the Dirichlet sum
\begin{equation}
\zeta(s) := \sum_{n=1}^\infty n^{-s}\;.
\label{zeta}
\end{equation}
By the fundamental theorem of arithmetic, this is equivalent to the Euler product over primes,
\begin{equation}
\zeta(s) = \prod_p(1-p^{-s})^{-1}\;.  
\label{zetap}
\end{equation}
The completed zeta function can be meromorphically continued to the full $s$ plane 
(with simple poles at $s=0$ and $s=1$) via Riemann's functional equation
\begin{equation}
\Lambda(1-s)=\Lambda(s)\;.
\label{func}
\end{equation}
The zeros of $\Lambda(s)$ coincide with the nontrivial zeros of $\zeta(s)$.

Berry and Keating \cite{BK} conjectured that the hamiltonian operator of the Hilbert--P\'olya conjecture should take the form
\begin{equation}
\Hbk := \half\omega_0(\x\p+\p\x)\;,
\label{Hbk}
\end{equation}
where $\x$ and $\p$ are position and momentum operators 
acting on some to-be-specificed compactification of the phase space for one degree of freedom.
(The parameter $\omega_0$ has dimensions of inverse time, and serves only to set the overall
energy scale.) 
Berry and Keating amassed numerous pieces of evidence in support of their conjecture.
For example, the Weyl formula for $N(E)$, the number eigenvalues of $\Hbk$ less that $E$, is
\begin{equation}
N(E)=\int \frac{dx\;dp}{2\pi\hbar}\,\theta(E-\omega_0 xp) + O(1)\;,
\label{ne}
\end{equation}
where $\theta(z)$ is the unit step function.
This integral diverges; to get a finite result, 
Berry and Keating impose lower cutoffs of $\ell_x$ and $\ell_p$ on the
integrations over $x$ and $p$, respectively, and take $\ell_x \ell_p = 2\pi\hbar$; then
they find \cite{BK}
\begin{equation}
N(E)=\frac{E}{2\pi\hbar\omega_0}\log\!\left(\frac{E}{2\pi\hbar\omega_0}\right)
-\frac{E}{2\pi\hbar\omega_0}+O(1)\;.
\label{ne2}
\end{equation}
This matches the asymptotic formula for the number of Riemann zeros with imaginary part between zero and $E/\hbar\omega_0$. 
With a further plausible assumption, the $O(1)$ term also agrees \cite{BK}. 

From here on, we set $\hbar\omega_0=1$ and (following a standard math convention) $h=2\pi\hbar=1$.  

Another intriguing aspect of $\Hbk$ is that its eigenfunctions (which have a twofold degeneracy) resemble the summand in the Dirichlet sum for the zeta function, with $s=\half-iE$ \cite{BK}.  
Writing the eigenvalue equation as
\begin{equation}
\Hbk|{E,}\delta\ra = E|{E,}\delta\ra\;,
\label{HbkE} 
\end{equation}
the (unnormalized) eigenfunctions are
\begin{equation}
\chi_{E,\delta}(x) := \la x|{E,}\delta\ra = (\sgn x)^\delta \, |x|^{-1/2+iE}\;,
\label{chiE}
\end{equation}
where $E$ is any real number and $\delta=0$ or $1$.
It seems like it ought to be possible to impose some sort of boundary condition on these eigenfunctions that ultimately
leads to the vanishing of $\Lambda(\half-iE)$ as an eigenvalue condition \cite{BK}.  
However, various attempts to carry out this
program have been unsuccessful to date \cite{others}.

In this paper, we will make a definite connection between the Berry--Keating hamiltonian and the
{\it local\/} Riemann hypothesis \cite{BN,BCKV,K}.

To explain the local Riemann hypothesis (which is actually a theorem \cite{BN,BCKV,K}), 
we first recall that Tate \cite{Tate} famously demonstrated that the structure of 
\eqs{Lam} and \eqref{zetap} is connected to the theory of valuations (norms) on
the field of rational numbers $\Q$.  In particular,
there is a valuation on $\Q$ associated with each prime number.
Given a prime $p$, any nonzero rational number $r$ can be written as
$r=\pm p^k a/b$, where $a$ and $b$ are positive integers not divisible by $p$, 
and $k$ is an integer (positive, negative, or zero);
the $p$-adic norm of $r$ is $|r|_p:=p^{-k}$.
Cauchy completion of $\Q$ with respect to this norm yields the $p$-adic numbers $\Qp$.  
The valuation on $\Q$ provided by the usual absolute-value norm
is not equivalent to any of these $p$-adic valuations, and instead is associated with a putative ``infinite prime''; we therefore denote the usual absolute value of $r$ as $|r|_\infty:=+p^k a/b$.
Cauchy completion of $\Q$ with respect to this norm yields the real numbers $\R=\Qinf$.
(For an introduction to $p$-adic numbers and valuations, see e.g.~\cite{Kob,VVZ}.)  

Let $x$ take values in $\Qv$ for some $v=\infty,2,3,\ldots$, 
and let $\psi_{v,0}(x)$ be the simplest complex-valued function of $x$ that is its own Fourier transform.
In particular, for $v=\infty$ (and hence $x\in\R$),
\begin{equation}
\psi_{\infty,0}(x):=e^{-\pi x^2}\;,
\label{psi0}
\end{equation}
where the Fourier transform over $\R$ is defined as
\begin{equation}
\tilde f_\infty(x):=\int_{-\infty}^{+\infty} dx\;e^{-2\pi i x y}f_\infty(y)\;.
\label{FT}
\end{equation}
For $v=p$ (and hence $x\in\Qp$),
\begin{equation}
\psi_{p,0}(x) :=
\begin{cases}
1 & \text{if } |x|_p\le 1\;, \\
0 & \text{if } |x|_p>1\;,
\end{cases}
\label{psi0p}
\end{equation}
where the Fourier transform over $\Qp$ is defined as
\begin{equation}
\tilde f_p(x):=\int_{\Qp}dx\;e^{2\pi i x y}f_p(y)\;;
\label{FTp}
\end{equation}
here $dx$ is the additive Haar measure on $\Qp$ normalized by $\int_{|x|_p\le1}dx=1$.
(For a review of $p$-adic calculus, see e.g.~\cite{VVZ}.)
We now define a ``local zeta function'' or ``gamma factor'' for each of the valuations,
\begin{equation}
\Gamma_{v,0}(s) := \int_\Qvx\dxx\;\psi_{v,0}(x)|x|_v^s\;,
\label{Gv0} 
\end{equation}
where the Haar measure $\dxx$ on the multiplicative group $\Qvx=\Qv\setminus\{0\}$ 
is given by $\dxx=\frac{v}{v-1}|x|_v^{-1}dx$.  For $v=\infty$, \eq{Gv0} evaluates to 
\begin{equation}
\Gamma_{\infty,0}(s) = \pi^{-s/2}\Gamma(\halfs)\;,
\label{Gi0} 
\end{equation}
which is precisely the completing factor in \eq{Lam}.  For $v=p$, \eq{Gv0} evaluates to
\begin{equation}
\Gamma_{p,0}(s) = (1-p^{-s})^{-1}\;,
\label{Gp0}
\end{equation}
which is precisely the Euler factor in \eq{zetap}.\footnote{Note, however, that \eq{Gi0} is not the $p\to\infty$ limit
of \eq{Gp0}; the relation of the infinite prime to the finite primes is more subtle.}  
Thus the completed zeta function can be written as a product of the gamma factors
over all the valuations or ``places'',
\begin{equation}
\Lambda(s) = \prod_{v=\infty,p}\Gamma_{v,0}(s)\;.
\label{Lamv}
\end{equation}

Next we note that the function $\psi_{v,0}(x)$ that appears in \eq{Gv0} is also the 
ground-state wave function of the harmonic oscillator 
(with the position and momentum both taking values in $\Qv$).   
The {\it local Riemann hypothesis\/} states that replacing the ground-state wave function 
$\psi_{v,0}(x)$ with an excited-state wave function $\psi_{v,N}(x)$ 
results in a modified gamma factor $\Gamma_{v,N}(s)$ that 
either vanishes identically, or whose zeros (if any) have real part one half.
This was proven for $v=\infty$ in \cite{BN,BCKV} and for $v=p\ne2$ in \cite{K}.  
(The case of $p=2$ remains unproven.)  Therefore, if the Riemann hypothesis is true,
then it is also true for the modified completed zeta function
\begin{equation}
\Lambda_{\{N_v\}}(s) := \prod_{v=\infty,p} \Gamma_{v,N_v}(s)\;,
\label{Lammod}
\end{equation}
with $N_v=0$ for all but a finite number of the places.  
Because only a finite number of the gamma factors are modified, 
the zeros of $\Lambda_{\{N_v\}}(s)$ are (after analytic continuation to the full $s$ plane)
the same as the Riemann zeros of the original $\Lambda(s)$, plus at most a finite number of
new zeros from the individual modified gamma factors.  

In this paper, we give a new proof of the local Riemann hypothesis for $v=\infty$, 
by showing that the modified gamma factor 
based on the oscillator eigenfunction $\psi_{\infty,N}(x)$ 
can be written as the spectral determinant of the Berry--Keating hamiltonian 
truncated to the subspace of oscillator eigenfunctions with index less than $N$.  
This is done in section \ref{Real}.
This demonstrates both the applicability of the Hilbert--P\'olya spectral approach in this simple model of the Riemann hypothesis,
and the relevance of the Berry--Keating hamiltonian to that approach.  
In section \ref{Padic}, we explore the $p$-adic case; we find that a spectral proof is possible,
but the relevant operator does not appear to be simply related to the Berry--Keating operator.
Conclusions are in section \ref{Conc}.

\section{The Local Riemann Hypothesis for $\R$}
\label{Real}

We want to show that the modified gamma factor for $v=\infty$, which we can write in a more prosaic form as 
\begin{equation}
\Gamma_{\infty,N}(s)=2\int_0^{\infty}dx\;\psi_{\infty,N}(x)x^{s-1}\;,
\label{GN}
\end{equation}
vanishes only if $\Re s=\half$.  
We begin by considering the inner product $\la N|{E,}\delta\ra$, 
where $|{E,}\delta\ra$ is an eigenstate of $\Hbk$, and $|N\ra$ is an eigenstate of the oscillator hamiltonian
$\Hho=\half(\x^2+\p^2)$ with eigenvalue $(2\pi)^{-1}(N+\half)$ ($N=0,1,\ldots$) and eigenfunction
\begin{equation}
\psi_{\infty,N}(x) := \la x|N\ra 
= \kappa_N H_N(\sqrt{2\pi}x)e^{-\pi x^2}\;.
\label{psiN} 
\end{equation}
Here $\kappa_N$ is a normalization constant and $H_N(y)$ is a Hermite polynomial.  
We note that $\psi_{\infty,N}(x)$ is real, has the symmetry property
\begin{equation}
\psi_{\infty,N}(-x)=(-1)^N\psi_{\infty,N}(x)\;,
\label{psi-x}
\end{equation}
and is its own Fourier transform, up to a fourth root of unity:
$\tilde\psi_{\infty,N}(x)=(-i)^N\psi_{\infty,N}(x)$.

Inserting a complete set of $\x$ eigenstates in $\la N|{E,}\delta\ra$, 
and using \eqs{chiE}, (\ref{psiN}), and the reality of $\psi_{\infty,N}(x)$, 
we have
\begin{align}
\la N|{E,}\delta\ra &= \int_{-\infty}^{+\infty}dx\;\la N|x\ra\la x|{E,}\delta\ra
\nonumber \\
&= \int_{-\infty}^{+\infty}dx\;\psi^*_{\infty,N}(x)\chi_{E,\delta}(x)
\nonumber \\
&= \int_{-\infty}^{+\infty}dx\;\psi_{\infty,N}(x)(\sgn x)^\delta|x|^{-1/2+iE}
\nonumber \\
&=
\begin{cases}
\Gamma_{\infty,N}(\half+iE) & \text{if }\delta = N\;\text{mod}\;2\;, \\
0 &\text{if }\delta \ne N\;\text{mod}\;2\;,
\end{cases}
\label{laNEra}
\end{align}
where the last line follows from \eqs{GN} and (\ref{psi-x}).

From here on we fix $\delta=N\;{\rm mod}\;2$.  We have shown that 
$\Gamma_{\infty,N}(\half+iE)=\la N|{E,}\delta\ra$.  The $\Hbk$ eigenstate $|{E,}\delta\ra$ is only
well-defined for real $E$, but the amplitude $\la N|{E,}\delta\ra$ can be analytically continued
to complex $E$.  We want to show that $\la N|{E,}\delta\ra$ vanishes only if $E$ is real.  

The condition $\la N|{E,}\delta\ra=0$ can considered as a kind of ``quantum boundary condition"
on the $\Hbk$ eigenstate $|{E,}\delta\ra$.  This terminology is motivated by the following analogy:
a Dirichlet boundary condition that a wave function $\phi(x)=\la x|\phi\ra$ vanish at some point $x=a$
can be written as the vanishing of the inner product $\la a|\phi\ra$, where $|a\ra$ is a position
eigenstate.  The quantum boundary condition generalizes this to the vanishing of the inner product
of $|\phi\ra$ with a more general state.

Unfortunately, there is no general principle that says that the analytic continuation of $\la N|{E,}\delta\ra$ 
to complex $E$ can vanish only if $E$ is real.  In the present case, however, we can show that
\begin{equation}
\la N|{E,}\delta\ra = c_N\,{\det}_N(E-\Hbk)\,\Gamma_{\infty,\delta}(\half+iE)\;,
\label{NEG}
\end{equation}
where again $\delta=N\;{\rm mod}\;2$, $c_N$ is a normalization constant,
and ${\det}_N$ means the determinant restricted to the subspace
of either either even (if $N$ is even) or odd (if $N$ is odd) oscillator eigenfunctions 
$\psi_{\infty,n}(x)$ with $n<N$.   
This spectral determinant can vanish only if $E$ equals an eigenvalue of $\Hbk$ in the restricted subspace.  
The restricted $\Hbk$ is hermitian, and so all its eigenvalues are real.  Since
$\Gamma_{\infty,0}(s)=\pi^{-s/2}\Gamma(s/2)$ and 
$\Gamma_{\infty,1}(s)=2\pi^{-s/2}\Gamma((s{+}1)/2)$ 
have no zeros, $\la N|{E,}\delta\ra$ vanishes
only if $E$ is a (real) eigenvalue of the restricted $\Hbk$.  This proves the local Riemann hypothesis
(previously proven by other methods in \cite{BN,BCKV}) for the real number field.  

To derive \eq{NEG}, we first define the usual oscillator raising and lowering operators 
$\a = \pi^{1/2}(\x + i\p)$ and $\ad = \pi^{1/2}(\x - i\p)$, which obey $[\a,\ad]=1$ and yield
$\Hho = (2\pi)^{-1}(\ad\a+\half)$.  In terms of $\a$ and $\ad$, we have
\begin{equation}
\Hbk =\half i(\ad\ad-\a\a)\;.
\label{Haa}
\end{equation}
Matrix elements of $\Hbk$ between eigenstates of $\Hho$ take the form
\begin{equation}
\la n|\Hbk|n'\ra =\half i \bigl[\sqrt{(n'+1)(n'+2)}\,\delta_{n,n'+2}-\sqrt{n'(n'-1)}\,\delta_{n,n'-2}\bigr]\;,
\label{nHbkn}
\end{equation}
where $\ad\a|n\ra = n|n\ra$ and $\la n|n'\ra=\delta_{n,n'}$.
We see that $n$ and $n'$ must be both even or both odd for the matrix element
to be nonvanishing, and so we can work in a subspace of either even or odd eigenstates of $\Hho$. 

We can treat both cases at once by setting $N=2K+\delta$, $n=2k+\delta$,
etc., where again $\delta=0$ if $N$ is even and $\delta=1$ if $N$ is odd.  
Then \eq{nHbkn} can be written as
\begin{align}
\la k|\Hbk|k'\ra &= b^*_{k'+1}\,\delta_{k,k'+1}+b_{k'}\,\delta_{k,k'-1}
\nonumber \\
&= b^*_{k}\,\delta_{k-1,k'}+b_{k+1}\,\delta_{k+1,k'} \;,
\label{kHbkk}
\end{align}
where $\ad\a|k\ra=(2k+\delta)|k\ra$, and $b_k=-(i/2)[(2k+\delta)(2k+\delta-1)]^{1/2}$.  
Note that $b_k\ne 0$ for $k>0$.

Consider $\la k|\Hbk|{E,}\delta\ra$.  We can evaluate this by letting $\Hbk$ act to the right, 
yielding its eigenvalue $E$.  Or, we can insert a complete set of eigenstates of $\Hho$ between
$\Hbk$ and $|{E,}\delta\ra$.  Equating the two results, we get
\begin{align}
E\,\la k|{E,}\delta\ra &= \sum_{k'=0}^\infty\la k|\Hbk|k'\ra\la k'|{E,}\delta\ra
\nonumber \\
&= b^*_{k}\la k{-}1|{E,}\delta\ra + b_{k+1}\la k{+}1|{E,}\delta\ra \;.
\label{ekHbke}
\end{align}
If we define wave functions
\begin{equation}
c_k\phi_k(E):=\la k|{E,}\delta\ra\;,
\label{chikE}
\end{equation}
where $c_k$ is a normalization constant to be fixed later, 
\eq{ekHbke} becomes
\begin{equation}
Ec_k\phi_k(E)=b^*_{k}c_{k-1}\phi_{k-1}(E)+ b_{k+1}c_{k+1}\phi_{k+1}(E) \;.
\label{EchikE}
\end{equation}
Rearranging and dividing through by $c_k$, we get
\begin{equation}
\frac{b_{k+1}c_{k+1}}{c_k}\phi_{k+1}(E)=E\phi_k(E)-\frac{b^*_{k}c_{k-1}}{c_k}\phi_{k-1}(E) \;.
\label{k1chikE}
\end{equation}
We now set $c_0=1$ and $c_{k+1}=c_k/b_{k+1}$.  \Eq{k1chikE} then simplifies to
\begin{equation}
\phi_{k+1}(E)=E\phi_k(E)-|b_{k}|^2\phi_{k-1}(E) \;.
\label{k1chikE2}
\end{equation}
The starting point for this recursion relation is 
$\phi_{-1}(E)=0$ and $\phi_0(E)=\Gamma_{\infty,\delta}(\half+iE)$.
The solution to \eq{k1chikE2} can then be written as
\begin{equation}
\phi_K(E) = 
\begin{vmatrix}
  E         &  -b_1     & \phantom{\ddots}   & {}        & {}               & {}           \\
 -b_1^*  &  E        & -b_2    &  \phantom{\ddots}      & {}                  & {}           \\
{}            & -b_2^*  & E       & \ \ddots     & {}              & {}           \\
{}            & {}            & \ddots  & \ \ddots  & \ddots       & {}           \\
{}            & {}            & {}         & \ \ddots  & E               & -b_{K-1} \\
{}            & {}            & {}         &  \phantom{\ddots}        & -b_{K-1}^* & E 
\end{vmatrix}
\phi_0(E) \;.
\label{phiKE}
\end{equation}         
To see that \eq{phiKE} is a solution to \eq{k1chikE2}, compute the determinant in \eq{phiKE} by minors on the last row.  \Eq{phiKE} is equivalent to \eq{NEG}, so we are done.

\section{The Local Riemann Hypothesis for $\Qp$}
\label{Padic}

The first issue to be faced is an appropriate definition of the Berry--Keating operator, and the first problem
is that derivatives of complex functions of a $p$-adic variable do not exist.  Therefore, in $p$-adic quantum
mechanics in general, it is necessary to define (instead of a hamiltonian) a unitary time-evolution operator for
a $p$-adic time variable $t$ \cite{VVZ}. 

In the real case, this time-evolution operator is simply $\exp(-i\Hbk t)$, and its position-space matrix
elements are 
\begin{equation}
\la x|\exp(-i\Hbk t)|y\ra = |e^{-\omega_0 t}|^{1/2}\,\delta(e^{-\omega_0 t}x-y)\;, 
\quad x,y,t,\omega_0 \in \R\;,
\label{xUy}
\end{equation}
where $\delta$ is the Dirac delta distribution.  For $x,y,t,\omega_0\in\Qp$, 
the right-hand side of \eq{xUy} is well-defined if we take $|\cdot|$ to be the $p$-adic norm, 
and require $|\omega_0 t|_p<1$, which allows the exponential function to be defined by a 
$p$-adically convergent Taylor series.  

This last requirement is too limiting, however, and so we instead choose to 
parameterize the evolution operator with a variable $a\in\Qp$ that can be thought of 
as the $p$-adic analog of $e^{-\omega_0 t}$.  We then define a $p$-adic analog of the
Berry--Keating evolution operator, $\Ubk(a)$, via
\begin{equation}
\la x|\Ubk(a)|y\ra := |a|_p^{1/2}\,\delta(ax-y)\;, \quad x,y,a \in \Qp\;.
\label{xUay}
\end{equation}
$\Ubk(a)$ is unitary, and obeys the composition rule
\begin{equation}
\Ubk(a)\Ubk(b)=\Ubk(ab)\;.
\label{UUU}
\end{equation}
The eigenstates of $\Ubk(a)$ are specified by the eigenfunctions
\begin{equation}
\chi_{E,\nu}(x) := \la x|{E,}\nu\ra = \nu(x)|x|_p^{-1/2+iE}\;,
\label{chiEnu} 
\end{equation}
where $E\in\R$
and $\nu(x)\in{\mathbb C}$ is a multiplicative character that obeys $|\nu(x)|_\infty=1$ and
$\nu(x)\nu(y)=\nu(xy)$.\footnote{If we set $x=p^k\xt$ with $|\xt|_p=1$,
then we have $\nu(x)=\nu(p)^k\nu(\xt)$.  Since $|x|_p=p^{-k}$, the phase factor of $\nu(p)^k$ 
in \eq{chiEnu} could be absorbed by shifting $E$ by a $p$-dependent constant.
However, it will be convenient to retain the redundant form of \eq{chiEnu}.} 
The corresponding eigenvalue of $\Ubk(a)$ is given by
\begin{equation}
\Ubk(a)|{E,}\nu\ra = \nu(a)|a|_p^{iE}\,|{E,}\nu\ra\;.
\label{Upsi} 
\end{equation}
We can now define a generalized gamma factor for $\Qp$ by
\begin{equation}
\Gamma_{p,0,\nu}(s) := \int_{\Qpx} \dxx\;\psi_{p,0}(x)\nu(x)|x|_p^s\;,
\label{Gnup} 
\end{equation}
where $\psi_{p,0}(x)$ is given by \eq{psi0p}.  

If we take $\nu(x)=1$ when $|x|_p=1$, then \eq{Gnup} evaluates to
\begin{equation}
\Gamma_{p,0,\nu}(s) = \frac{1}{1-\nu(p)p^{-s}}\;,
\label{Gnup2} 
\end{equation}
and the product over $p$ yields the Dirichlet $L$ function
\begin{equation}
L(s,\nu) := \sum_{n=1}^\infty \frac{\nu(n)}{n^s} = \prod_p \Gamma_{p,0,\nu}(s) \;.
\label{Lsnu}
\end{equation}
Allowing for appropriate nontrivial values of $\nu(x)$ when $|x|_p=1$
yields extra factors that, when
combined with the gamma factor for $v=\infty$, complete the $L$ function into one
that obeys a simple functional equation.  

From here on, we assume that $p\ne2$.

Next we must analyze the $p$-adic generalization of the harmonic oscillator  \cite{VVZ}.  
This is specified by a unitary evolution operator $\Uho(c,s)$ with position-space matrix elements
\begin{equation}
\la x|\Uho(c,s)|y\ra
= \kappa(c,s)|s|_p^{-1/2}p^{n/2}\exp[2\pi i(c x^2 - 2 x y + c y^2)/sp^n]\;,
\label{Uho}
\end{equation}
where $c$ and $s$ are $p$-adic parameters that obey $c^2+s^2=1$, 
$\kappa(c,s)$ is an eighth root of unity, and $p^n$ ($n\in\Z$) is a normalization constant.
In terms of a $p$-adic time parameter $t$, we would have 
$c=\cos t$ and $s=\sin t$, but this overly restricts us to $|c|_p=1$ and $|s|_p<1$.
Also, by rescaling both $x$ and $y$ by a power of $p$, we can change $n$ to
either 0 or 1, depending on whether $n$ is even or odd.  From now on we specialize
to the case $n=0$; the case $n=1$ is treated similarly.  

The oscillator ground state $|0\ra$ obeys $\Uho(c,s)|0\ra =|0\ra$, and is specified by the
wave function $\la x|0\ra = \psi_{p,0}(x)$, given by \eq{psi0p}.  We note that
\begin{align}
\la 0|{E,}\nu\ra &= \int_{\Qp} dx\;\la 0|x\ra\la x|{E,}\nu\ra
\nonumber \\
&= \int_{\Qp} dx\;\psi^*_{p,0}(x)\chi_{E,\nu}(x)
\nonumber \\
&= \int_{\Qp} dx\;\psi_{p,0}(x)\nu(x)|x|_p^{-1/2+iE}
\nonumber \\
&= \frac{p-1}{p}\int_{\Qp} \dxx\;\psi_{p,0}(x)\nu(x)|x|_p^{+1/2+iE}
\nonumber \\
&= \frac{p-1}{p}\,\Gamma_{p,0,\nu}(\half+iE)\;.
\label{<0E>}
\end{align}
So, as in the real case, the $p$-adic gamma factor is given by the inner product
of oscillator ground state with an eigenstate of the Berry--Keating operator.  
We wish to consider modifying the gamma factor by replacing the oscillator
ground state with an excited state.

Let $|N\ra$ denote an excited state of the $p$-adic harmonic oscillator,
\begin{equation}
\Uho(c,s)|N\ra = \gamma_N(c,s)|N\ra\;,
\label{UhoN} 
\end{equation}
where $|\gamma_N(c,s)|_\infty=1$.  Excited states are always degenerate
(infinitely so if $p=1\mathrel{\rm mod}4$); 
they are partially specified by a positive integer $N$.
The eigenfunction $\psi_{p,N}(x)=\la x|N\ra$ vanishes if $|x|_p>p^N$, and is locally constant:
$\psi_{p,N}(y)=\psi_{p,N}(x)$ if $|x-y|_p\le p^{-N}$.
Kurlberg \cite{K} has shown that an eigenfunction for a given value of $N$
takes the form
\begin{equation}
\psi_{p,N}(x) = \Delta_N(x)+\lambda_N\tilde\Delta_N(x) + f_N(x)\;,
\label{psiNx}
\end{equation}
where $\lambda_N=\pm 1$, 
\begin{equation}
\Delta_N(x) :=
\begin{cases}
1 & \text{if } |x|_p=p^N\;, \\
0 & \text{otherwise}\;,
\end{cases}
\label{dN}
\end{equation}
$\tilde\Delta_N(x)$ is the Fourier transform of $\Delta_N(x)$, given explicitly by
\begin{equation}
\tilde\Delta_N(x) = p^N\Omega_{-N}(x)-p^{N-1}\Omega_{-N+1}(x)\;,
\label{tpsiNx}
\end{equation}
where
\begin{equation}
\Omega_N(x) :=
\begin{cases}
1 & \text{if } |x|_p \le p^N\;, \\
0 & \text{if } |x|_p > p^N\;,
\end{cases}
\label{thN}
\end{equation}
and $f_N(x)=\tilde f_N(x)$ is real and obeys
\begin{equation}
\int_{|x|=p^k}dx\;f_N(x)=0 \quad\forall k\in{\Z}\;.
\label{intpk}
\end{equation}

From here on, for simplicity, we specialize to the case that $\nu$ is trivial, $\nu(x)\equiv 1$,
and drop the subscript $\nu$.
The case of nontrivial $\nu$ is treated by Kurlberg \cite{K}.

Mimicking \eq{<0E>}, we have
\begin{equation}
\frac{p}{p-1}\la N|{E,}1\ra = \int_{\Qp} \dxx\;\psi_{p,N}(x)|x|_p^{+1/2+iE}\;.
\label{<NE>}
\end{equation}
Setting $s=\half+iE$, and using eqs.$\,$(\ref{psiNx}--\ref{intpk}), the result is
\begin{equation}
\frac{p}{p-1}\la N|{E,}1\ra = p^{Ns}
+ \frac{\lambda_N}{1-p^{-s}}\bigl(p^N p^{-Ns} - p^{N-1}p^{-(N-1)s}\bigr)\;.
\label{<NE>2}
\end{equation}
Following Kurlberg \cite{K}, this can be simplified by setting $\alpha=p^{-1/2}$ and $y=p^{iE}$,
which yields
\begin{equation}
\frac{p}{p-1}\la N|{E,}1\ra 
= \frac{y^{2N}-\alpha y^{2N-1}-\lambda\alpha y + \lambda}{\alpha^N y^{N-1}(y-\alpha)}\;,
\label{<NE>3}
\end{equation}
where we have dropped the subscript $N$ on $\lambda$ to streamline the notation.
For finite $E$ (and hence finite $y$), this expression can vanish only if the numerator vanishes.
We now note that the numerator can be written as the spectral determinant of 
a relatively simple $2N\times 2N$ unitary matrix.  Specifically,
\begin{equation}
y^{2N}-\alpha y^{2N-1}-\lambda\alpha y + \lambda = \det(y - U)\;,
\label{ydet}
\end{equation}
where 
\begin{equation}
U =
\begin{pmatrix}
\alpha                 &  \beta                   & \phantom{\ddots}        & {}        & {}                        \\
{}                & \ 0  \                   & \ 1 \       & \phantom{\ddots}         & {}                          \\
{}                 & {}                            & \ddots  & \ddots       & {}           \\
{}                 & {}                          & \phantom{\ddots}         & 0              & \ 1 \       \\
-\lambda\beta  & \lambda\alpha  & \phantom{\ddots}             & {}               &   0 
\end{pmatrix}
\label{U}
\end{equation}  
with $\alpha=p^{-1/2}$ and $\beta=(1-\alpha^2)^{-1/2}=(1-p^{-1})^{-1/2}$.        
Since $U$ is unitary, $y=p^{iE}$ must have unit magnitude if $\det(y-U)$ vanishes, and so $E$ must be real.

Unfortunately, though $U$ is a relatively simple matrix, it does not appear to be possible to regard it as 
a truncation of the Berry--Keating operator $\Ubk(a)$ for some value of $a$.  
Thus a spectral proof of the local Riemann hypothesis in the $p$-adic case 
appears to require construction of a more complicated unitary operator.  
A more detailed investigation of $p$-adic oscillator eigenfunctions would be needed
to construct this operator on the full Hilbert space. 

\section{Conclusions}
\label{Conc}
We have investigated the local Riemann hypothesis from the point of view of the 
Hilbert--P\'olya conjecture.  In the case of the real number field, we have given a new proof
of the local Riemann hypothesis, and shown that the imaginary parts of the zeros are the
eigenvalues of the Berry--Keating operator restricted to a finite-dimensional subspace.
The situation in the $p$-adic case is more complicated, with the zeros related to the eigenvalues
of a unitary matrix that does not appear to be simply related to the $p$-adic generalization
of the Berry--Keating operator.  Further investigation is needed to see
if this unitary matrix can be given a natural interpretation.
\acknowledgments
I thank P\"ar Kurlberg and Michael Berry for helpful correspondence.  
This work was supported in part by the National Science Foundation under grant PHY07-57035.

\end{document}